\documentclass[conference]{IEEEtran}
 
\usepackage{cite}

\ifCLASSINFOpdf
   \usepackage[pdftex]{graphicx}
\else
   \usepackage[dvips]{graphicx}
\fi
\usepackage{pgfplots}
\usepackage{pgfplotstable}
\usepackage{booktabs}
\usepackage{array}
\usepackage{colortbl}
\pgfplotstableset{
	every head row/.style={before row=\toprule,after row=\midrule},
	every last row/.style={after row=\bottomrule},
	fixed,precision=2,
}

\usepackage{amsmath}
\usepackage{array}

\usepackage{url}

\usepackage[]{fancyhdr} %
\newcommand{\changefont}{\fontsize{9}{9}\selectfont}
\IEEEoverridecommandlockouts
\begin{document}

%
\title{Proposal and Description of a Test System with Wind, Hydro and Fossil Fuel Power Plants for Static Analyses}

\author{\IEEEauthorblockN{Line 1: Authors Name/s per 1st Affiliation\\Line 2: Author's Name/s per 1st Affiliation}
\IEEEauthorblockA{Line 3 (\textit{of Affiliation}): Dept. name of organization\\Line 4: name of organization, acronyms acceptable\\
Line 5: City, Country\\
Line 6: e-mail address if desired}
\and
\IEEEauthorblockN{Line 1: Authors Name/s per 2nd Affiliation\\Line 2: Author's Name/s per 1st Affiliation}
\IEEEauthorblockA{Line 3 (\textit{of Affiliation}): Dept. name of organization\\Line 4: name of organization, acronyms acceptable\\
Line 5: City, Country\\
Line 6: e-mail address if desired} 

}


%

	\author{\IEEEauthorblockN{Victor Neumann\IEEEauthorrefmark{1}\IEEEauthorrefmark{3},
		Roman Kuiava\IEEEauthorrefmark{2},
		Rodrigo A. Ramos\IEEEauthorrefmark{3}  and
		Ahda Pavani\IEEEauthorrefmark{4}}
	\IEEEauthorblockA{\IEEEauthorrefmark{1} Department of Engineering and Exact Sciences - 
		Federal University of Parana,
		Palotina - PR - Brazil - 85950-000\\ vneumann@ufpr.br, vneumann@usp.br}
	\IEEEauthorblockA{\IEEEauthorrefmark{2} Department of Electrical Engineering - 
		Federal University of Parana,
		Curitiba - PR - Brazil - 81530-000\\ kuiava@ufpr.br}
	\IEEEauthorblockA{\IEEEauthorrefmark{3} Department of Electrical Engineering -
		Engineering School of Sao Carlos - USP,
		Sao Carlos - SP - Brazil - 13566-590\\ rodrigo.ramos@ieee.org}
	\IEEEauthorblockA{\IEEEauthorrefmark{4}Center for Engineering and Applied Social Sciences -
		Federal University of ABC,
		Santo Andre - SP - Brazil - 09210-580\\ ahda.pavani@ufabc.edu.br}
}


\lhead{In proceedings of the 11th Bulk Power Systems Dynamics and Control Symposium (IREP 2022), July 25-30, 2022, Banff, Canada}


\maketitle
\thispagestyle{fancy}
\pagestyle{fancy}


\begin{abstract}
	This article presents and describes a 229 bus test system that includes wind, hydro and fossil fuel power plants. It represents the Northeast subsystem of the Brazilian Interconnected Power System (BIPS). The test system supplies a load of 4.17\hspace{0.05cm}GW, being 13\% powered by wind farms, which is the current wind power penetration level of the BIPS. The data comprehends different load levels based on the typical load behavior and typical capacity factors of wind, hydro and fossil fuel plants, as well as the capacity of transmission and sub-transmission lines, transformers, and the adopted structure for the test system. The data is compiled considering models and operating scenarios of the BIPS, and allow performing studies of static voltage stability, sensitivity of voltage stability margin considering the wind farms, and multi-objective optimization considering market constraints. The results of the simulations with the test system indicate the consistency of their data structure and its applicability to different studies of electric power systems. 
\end{abstract}

\section{Introduction}	
At the beginning of the year 2022, the Brazilian Interconnected Power System (BIPS) surpassed 20 GW of wind power installed capacity, spread in 751 wind farms with a total of 8,800 wind turbines  \cite{ASH3:2021:Online}. Around 90\% of this total capacity is located at the Northeast (NE) subsystem of the BIPS. Due to the significant penetration level of wind power in this subsystem and the need to perform several studies considering wind generation and its interaction to the 
other power plants, this paper presents a 229-bus test system composed of wind, hydro, and fossil fuel power plants, named NE-BIPS-229bus.	
\par
\begin{figure}[h]
	\centering
	\includegraphics[width=1\linewidth]{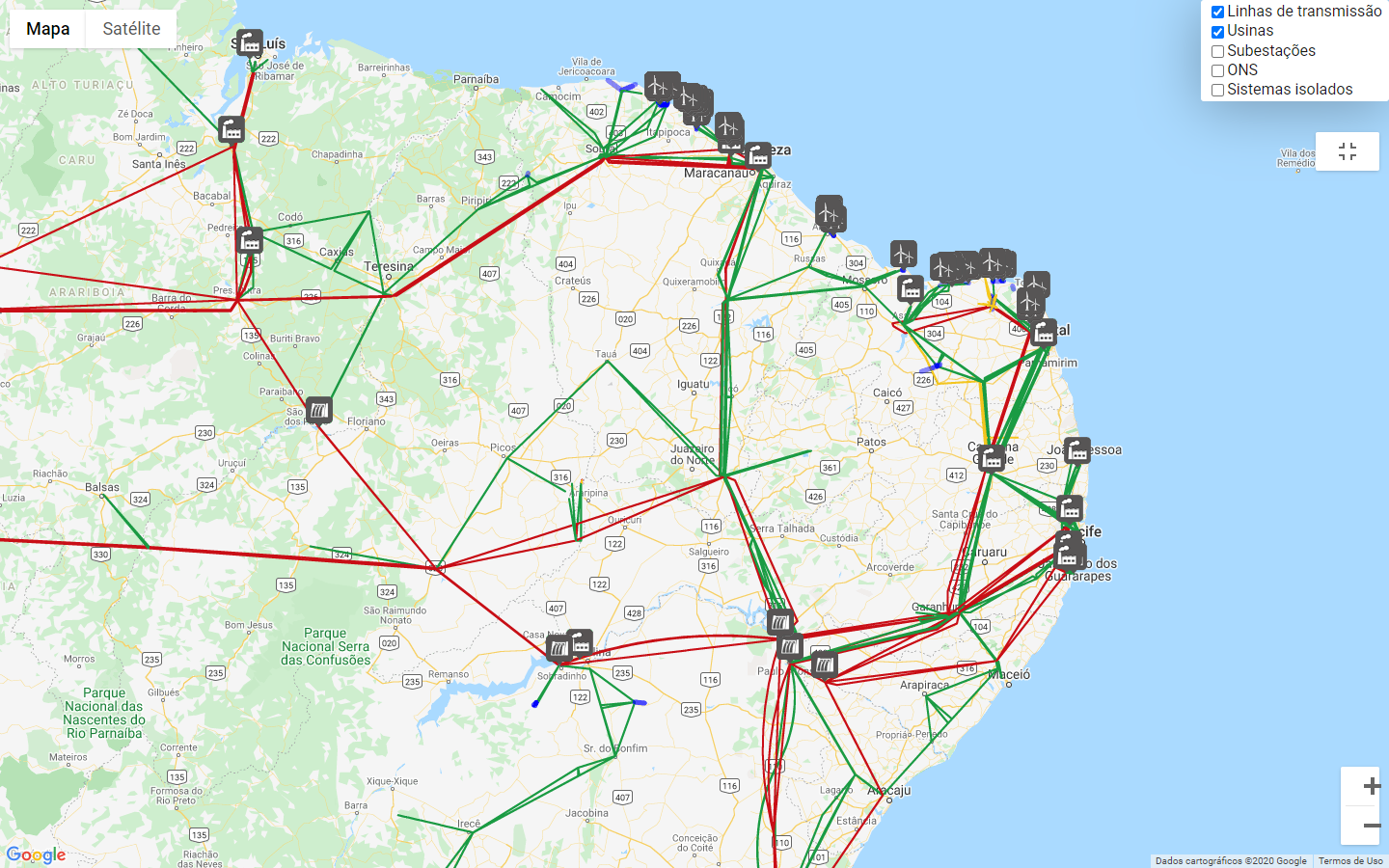}
	\caption{Coverage of the 229-bus test system - Source: BIPS Dynamic Map - ONS \cite{ASH6:2019:Online}}
	\label{fig:screenshot002}
\end{figure}

 The test system is based on the NE subsystem due to the projection of connection of more wind farms to this subsystem in short, mid, and long term. The region presents load centers and power plants in the states of Ceará (CE), Rio Grande do Norte (RN), Paraíba (PB), Pernambuco (PE), Alagoas (AL), Sergipe (SE) and Bahia (BA), covering an approximated area of 720,000 Km$^2$, whose geographic area is in Figure \ref{fig:screenshot002}. 
\par The test system contemplates a total load of 4.17\hspace{0.05cm}GW, representing the annual average load level of the NE subsystem for 2021, with 13\% of this load supplied by wind farms, which represents the current wind power penetration in the BIPS. Some electrical and operating data of the test-system are: a total load of 4.17\hspace{0.05cm}GW, spread in 42 load buses; total losses of 85 MW; total generation of 4.25\hspace{0.05cm}GW; 262 transmission and subtransmission lines, with voltage levels of 500, 230, 138, 69, 34.5, and 20\hspace{0.05cm}kV; and 159 transformers. The load is powered by 31 wind farms, 19 hydro power plants and 18 fossil fuel power plants, corresponding to, respectively, 13\%, 58\% and 29\% of the total generation capacity. This mix of synchronous and assynchronous generation represents the main feature of the proposed test system, since it allows the analyzes of the main issues that may arise in power systems in the transition to an increase of the penetration levels of renewable sources connected to the grid by inverters.
\par The Figure \ref{fig:screenshot002} shows the geographic coverage of the test system in the NE region of Brazil, which illustrates the locations of the main wind farms, hydro and fossil fuel power plants. Also, the main power flow corridors in the 500 KV (in red) and 230 KV (in green) transmission lines are illustrated. 
\par The precise geographical location and details of each power plant, transmission system and substations can be obtained from the BIPS's Geographical Information System Dynamic Map \cite{ASH6:2019:Online}.
\subsection{Literature review}
There are numerous test systems available at different sites that can be used to perform various types of simulations aimed at studying electrical power systems, among which we can highlight the widely used IEEE test systems from 14 to 300 buses available on the University of Washington website \cite{ASH1:2022:Online}. An update of the IEEE 118-bus, named NREL-118, with a reconfigured generation representation using three regions of the US Western Interconnection from the latest Western Electricity Coordination Council (WECC) 2024 is detailed in \cite{articleIEEE118up}. The NREL-118 data includes generation technologies with different heat functions, stable minimum levels and ramp rates, greenhouse gas (GHG) emissions rates, regulation and contingency reserves, solar and wind hourly time series provided for one year. 
\par Also, noteworthy are the test systems from 9 to 107 buses that are widely used in static analysis, and in some cases dynamic, of BIPS subsystems, available in \cite{ASH2:2022:Online}. There are also work that provide specific models to test algorithms of control and protection techniques in the integration of microgrids and generation from renewable sources \cite{7339813}.
\par However, as the test system described in this article has the specific objective of enabling the analysis of the main problems that may arise in energy systems in the transition to an increase of penetration levels of renewable sources connected to the grid by inverters, this research explored the available literature on studies of the impacts on the electrical system by the coupling of wind farms, and mainly on their interconnection points (POIs). These POIs can represent reactive power reservoirs when the renewable generation coupled to the system, mainly at the sub-transmission level, comes from wind or solar photovoltaic farms, and may lead to diverses consequences on system stability, as highlighted in \cite{7285805}. 
\par Though, \cite{6039797} mentions that the behavior of photovoltaic generation in the electric grid is determined by the way the active and reactive output power are controlled by the inverters, allowing the control of the voltage magnitude in the POI. Other studies point out that if the coupled distributed generation is from plants with synchronous generators, which may be from small fossil fuel and hydro power plants  generators, the control variables of these generators \cite{574947}, such as power factor (PF) adjustment \cite{hatziargyriou2017contribution}, could be used as resources that contribute to the stability of the electrical system, or at least in the vicinity of the POI. \par Other works have proposed the use of reactive power generation capacity of wind farms to improve the transient stability of the power system for the improvement of Fault-ride Through (FRT) to reduce losses of the system and to attenuate the voltage fluctuations \cite{{en9121066}, {meegahapola2013capability}, {5159366}, {5510193}}. Also, an enhanced reactive power capability based wind farms control strategy, during stressed voltage conditions to alleviate system instability, measured through maximum power transfer capability of the network is proposed in \cite{9126776}, and an active voltage control strategies, including active adjustment of reactive power reference, active speed control according to wind speed ranges, are proposed in \cite{8668559}.
\subsection{Contributions}
\par An important feature of the proposed test system is that it corresponds to a subsystem of a large-scale real power system and, therefore, includes issues that a real system is subjected to. The test system can be used in studies for planning the expansion and operation of power systems, considering different aspects and tools such as power flow analyses, voltage stability, sensitivity analysis for voltage stability margin, optimum power flow (considering voltage and generation restraints) for optimum dispatch, congestion management, allocation of renewable generation and shunt compensators. To make these studies and analyses feasible, transmission and sub-transmission systems, different load levels, generation systems and their capacity factors were modeled according to their primary energy source.	
\par Given this context, the paper describes the data structure of the subtransmission and transmission system, transformers, reactive power compensation devices, loads, and generation from wind, hydro, and fossil fuel  power plants for the NE-BIPS-229bus. These structures were compiled with detailed electrical data for the base case of the NE subsystem of BIPS.
\subsection{Paper organization}	
\par This paper is organized as follows: Section II discusses the model adopted to estimate the NE subsystem load levels, and describes the areas, regions, buses, and their data structures used in the test system. Section III comprehends the model of the generation buses and their data structures, as well as the capacity factors of wind, hydro, and fossil fuel power plants are presented. Section IV describes the data structure and the model parameters of the transmission system and transformers. Section V addresses the modeling of supply and generation cost, and demand prices for elastic loads that are part of the short-term market contracted in the free energy market (in the case of Brazil) or market bidding. Section VI presents the simulation results using the test system, which results allow verifying the applicability of the test system in different studies. Section VII highlights the main conclusions and the future steps to improve the proposed test system.    

\section{Model of the Load Levels, Areas, Regions, Buses, and their Data Structure}	
This section presents the adopted modelling procedure to estimate the load parameters for the subsystem, considering its monthly behavior. In addition, it is also presented the data structure of areas, regions, and load buses with constant power.

\par The modeling of part of the NE subsystem to structure the test system, proposed in this paper, adopts the class structure format (MatLab version) due to its versatility to structure data and manipulate them with MatLab or Octave  coding, among others.  This class structure was adopted by the Power System Analysis Toolbox (PSAT)  \cite{milano2006open}. This paper uses the PSAT since it is open source and it is able to recognize and convert a wide variety of test system formats usually employed in power system research. Additionally, the results presented in this paper were obtained with simulations using different tools available in PSAT. 

\par In the presentation of the data structure, the formulations for the use of these structure related to the active and reactive power equations to be used in the power balance are also presented. These formulations can be useful for the user who wants to write his own code to carry out simulations with this test system.  

\par The test system is available in \cite{ASH:20214:Online}. In this link, there is a version containing a commented version, the file NE-BIPS-229bus.m (MatLab); in version with purely numeric data structures, considering the average load level scenario for 2021 and capacity factors obtained for September of the same year, in the NE-BIPS-229bus.txt; and the one-phase diagram is available in 229bus-Single-Line-Diagram.pdf. The access can be required by sending an e-mail to one of the authors.

\subsection{Data structure of areas, regions, buses and their names}
Areas and Regions structures are used to geographically identify all the buses that belong to a specific subsystem of the BIPS. The Areas structure is used to identify the subsystem and the Regions structure to geographically identify the buses by state of the federation. Particularly for the NE subsystem, integers are assigned to each state in the northeast region of the country, with the exception of the state of Maranhão, which in the BIPS was established as belonging to the north.	
\par The Areas and Regions structures are important to verify the inter-area and inter-regions power exchange \cite{4334890}, besides it allows defining a reference bus to a specific area or region (it is assigned a null value when the region/area does not have a reference bus). Table~\ref{tabela 11} shows the data structure of Areas (Areas.con). The Regions data structure (Regions.con) presents the same data structure of areas. In the test system, all class structures and data have the title with file extension .con, for instance, Areas.con.	The data structure of bus, area, and region names are the same, as presented in Table~\ref{tabela 12}. 
\begin{table}[h]
	\renewcommand{\arraystretch}{1.1}
	\centering
	\caption{Areas data structure (Areas.con)}
	\label{tabela 11}
	\begin{tabular}{|c|c|l|c|}
		\hline 
		\textit{Column} & \textit{Variable} & \textit{Description} & \textit{Unit} \\
		\hline 
		1 & $\#$ & \textit{Area Number} & \textit{int} \\ 
		
		2 & $\#$ & \textit{Slack bus number for the area} & \textit{int} \\  
		
		3 & $S_b$ & \textit{Power base} & \textit{MVA} \\ 
		
		4 & $P_{ex}$ & \textit{Exported interchange ($>$ 0 = injecting)} & \textit{p.u.} \\
		
		5 & $P_{tol}$ & \textit{Interchange acceptable error} & \textit{p.u.} \\
		
		6 &$\Delta P_{\%}$ & \textit{Annual growth rate} & \textit{\%} \\
		\hline 
	\end{tabular} 
\end{table}

\begin{table}[h]
	\renewcommand{\arraystretch}{1.1}
	\centering
	\caption{Names data structure (Bus/Areas/Regions.names)}
	\label{tabela 12}
	\begin{tabular}{|c|c|c|c|}
		\hline 
		\textit{Column} & \textit{Variable}& \textit{Description} & \textit{Unit} \\ 
		\hline 
		1 & $\#$ &\textit{'Names of Bus / Area / Region'} & \textit{txt} \\ 
		\hline 
	\end{tabular}
\end{table} 
\par As an example, Table~\ref{tabela 13} presents the name of the 8 regions, which are related to the NE states of Brazil. As the structure has only one column, the table also includes the region names (\textit{txt}), which are related to the states. Similarly, the area names include the 4 areas (subsystems) and the names of buses include the nomenclatures of the 229 buses, as detailed in NE-BISP-229bus.	
\begin{table}[h]
	\renewcommand{\arraystretch}{1.1}
	\centering
	\caption{Regions names (Regions.names)}
	\label{tabela 13}
	\begin{tabular}{|l|}
		\hline
		\textit{Regions}\\
		\hline
		Alagoas (AL)\\ Bahia (BA)\\ Ceará (CE)\\  Paraíba (PB)\\ Piauí (PI)\\ Pernambuco (PE)\\ Rio Grande do Norte (RN) \\ Sergipe (SE)\\
		\hline
	\end{tabular}
\end{table} 
\begin{table*}[h]
	\renewcommand{\arraystretch}{1.1}
	\centering
	\caption{Data of centroid of heavy load level (LoadLev.Haevy.con)}
	\label{tabela 26}
	\begin{tabular}{|c|c|c|c|c|c|c|c|c|c|c|c|c|c|}
		\hline 
		\textit{Area number}&\textit{${HL}_1$} &\textit{${HL}_2$}& \textit{${HL}_3$}& \textit{${HL}_4$}& \textit{${HL}_5$} &\textit{${HL}_6$}& \textit{${HL}_7$}& \textit{${HL}_8$}&\textit{${HL}_9$}&\textit{${HL}_{10}$}& \textit{${HL}_{11}$}& \textit{${HL}_{12}$}\\
		\hline
		3&1.0990& 1.1009&	1.0905&	1.0968&	1.0959&	1.0923&	1.0899&	1.0939&	1.0990&	1.0936&	1.0972& 1.1050\\
		\hline
	\end{tabular}
\end{table*}	
\subsection{Model of load levels of the BIPS}\label{sec:modelagem-dos-patamares-de-cargas-do-sin}
Currently, the planning and operation of the BIPS represent the monthly load duration curves in three levels associated with fixed time bands, determined according to the day of the week \cite{ASH9:2017:Online}. These load levels are defined as heavy, medium and light. The BIPS load curve has been changing, with demand peaks occurring in the middle of the afternoon, especially in summer, and also by demand management actions, which reduce demand exactly during heavy load times.
\par The decoupling between the load curve and its representation in levels implies distortions in expansion and operation planning, in the formation of energy prices, in the contracting of generation and in the valuation of tariff stations. The study Representation of Load Levels in the Chain of Computational Models in the Electric Sector \cite{ASH8:2019:Online} reassesses the discretization of the load curve into levels, proposing the application of the Cluster Analysis methodology as a useful statistical technique for segmenting data sets to represent load levels.
\par The data used by the above-mentioned study methodology comes from the load history of the Southeast/Midwest, South, Northeast and North subsystems in the period from 01/01/2012 to 12/31/2017. Among the results presented in the work \cite{ASH8:2019:Online} and obtained by the Cluster Analysis, a table can be highlighted that includes the Distribution of Load Records by Levels, Duration of Levels and Centroid of Levels, for each month of the year. This proposal incorporates in the NE-BISP-229bus test system as heavy, medium and light loads, these centroids of load levels, which are referenced to the average annual load of each subsystem (Area) as 1.0 p.u. 
\subsection{Load levels data structure}
In order to have a uniform data structure for all load levels, Table~\ref{tabela 25} presents the structure that is common for the heavy (\textit{HL}), medium (\textit{ML}) and light (\textit{LL}) load levels. 
\begin{table}[h]
	\renewcommand{\arraystretch}{1.1}
	\centering
	\caption{Load levels data structure (LoadLev.Haevy/Medium/Light.con)}
	\label{tabela 25}
	\begin{tabular}{|c|c|l|c|}
		\hline
		\textit{Column} & \textit{Variable} & \textit{Description} & \textit{Unit} \\ 
		\hline 
		1 & \textit{${A}_i$} & \textit{Area number} & \textit{int} \\ 
		2 & \textit{${HL}_m$} / \textit{${ML}_m$} / \textit{${LL}_m$} & \textit{Load level} & \textit{real} \\ 
		\hline
	\end{tabular}
\end{table}
\par The first row of Table~\ref{tabela 25} includes the subsystem number, which is assumed as Area. In the second row, the variable $\textit{HL}$, $\textit{ML}$, and $\textit{LL}$ are vectors containing the centroids of the heavy, medium, and light load levels, respectively, for each month $m$. For example, Table~\ref{tabela 26} presents the centroids of heavy load levels $\textit{HL}$ of NE subsystem (\#\hspace{0.05cm}3) for the whole year. The data of centroids of medium and light load levels presents the same structure and are available in the NE-BIPS-229bus.
\par As this research had access to the real data of the average annual load of the NE subsystem of the BIPS for the year 2017, the estimation of average load for 2021 was performed according to the load growth forecast used in the Decennial Plan for Energy Expansion for the Brazilian system (PDE): 
PDE-2027, 3.7\% used for annual periods of 2017-2018-2019;	
PDE-2029, 3.6\% used for annual periods of 2019-2020; and	
PDE-2030,  1.8\% used for annual period of 2020-2021, predicting a reduction of the growth rate until 2025 due to Covid-19 pandemic
\cite{ASH:2021:Online}.	
Thus, the total adjustment applied to estimate the average load of NE subsystem for 2021 is 1.037*1.037*1.036*1.018\hspace{0.1cm}=\hspace{0.1cm}1.1341. This approach to estimate the load can be extended to any annual period of load growth prediction to be used by the Brazilian \textit{Empresa de Pesquisas Energéticas} (EPE) \cite{ASH:2021:Online}.	
\par The load data used in PQ buses are those estimated as annual average load to the year of 2021 for the NE subsystem. The simulation results using the test system, presented in section   \ref{sec:resultados-de-simulacoes}, were performed using the centroids of heavy-load level (1.0990), medium (0.9865), and light (0.8779) for September. 
The formulation that can be used to estimate bus loads, for exemple, for the heavy load level for a given month $m$ of the year is as follows: 
\begin{subequations} 
	\label{keyE6.11211} 
	\begin{align} 			
		P_{L_H} &= \textit{HL}_mP_{L} \label{eq:line1011}\\
		Q_{L_H} &= \textit{HL}_mQ_{L} \label{eq:line1012}		
	\end{align}
\end{subequations}
\noindent where $P_{L}$ e $Q_{L}$ are the vectors of active and reactive powers following the data structure presented in Table \ref{tabela 3}, which are considered as load of PQ buses of the NE-BIPS-229bus.	
\subsection{Bus data structure}	
The topology of the electrical grid is defined by the components of the bus data structure, and its components are represented in Table~\ref{tabela 1}. In this structure, the bus number $\#$ (integer) and the magnitude of the voltage base $V_b$ (kV) are mandatory. The other structure components are optional.
The bus name is associated to the bus number of Table~\ref{tabela 1} using the data structure of bus name (Bus.names) of Table~\ref{tabela 12}. The Bus.con class structure incorporates the Bus.names structure. The same procedure happens to other structures that include names, such as Areas and Regions.
\begin{table}[h]
	\renewcommand{\arraystretch}{1.1}
	\centering
	\caption{Bus data structure (Bus.con)}
	\label{tabela 1}
	\begin{tabular}{|c|c|l|c|}
		\hline
		\textit{Column} & \textit{Variable} & \textit{Description} & \textit{unit} \\ 
		\hline 
		\rule[-1ex]{0pt}{2.5ex} 1 & $\#$ & \textit{Bus number} & \textit{int} \\ 
		\rule[-1ex]{0pt}{2.5ex} 2 & $V_b$ & \textit{Voltage base} & \textit{kV} \\ 
		\rule[-1ex]{0pt}{2.5ex} 3 & $V_0$ & \textit{Voltage magnitude initial guess} & \textit{p.u.} \\ 
		\rule[-1ex]{0pt}{2.5ex} 4 & $\theta_0$ & \textit{Voltage phase initial guess} & \textit{rad} \\ 
		\rule[-1ex]{0pt}{2.5ex} 5 & $A_i$ & \textit{Area number} & \textit{int} \\ 
		\rule[-1ex]{0pt}{2.5ex} 6 & $R_i$ & \textit{Region number} & \textit{int} \\ 
		\hline
	\end{tabular}
\end{table}	
\subsection{PQ bus data structure}	
The loads of the electrical system are modeled as active powers $P_L$ and reactive powers $Q_L$ constants, and their data structure is given by Table~\ref{tabela 3}. All the nine entries in this structure should be filled, and the most important entries are the load active $P_L$ and reactive power $Q_L$. Each bus has its power base $S_b$, and voltage base $V_b$ (p.e. 34.5 kV is typical the terminal voltage at the point of connection of a wind farm). $V_{min}$ and $V_{max}$ are the voltage limits of each bus in p.u., which is usually defined as 0.9 and 1.1 p.u., respectively. In the case it is necessary that the voltage limits can be violated, the load can be converted to a constant impedance (setting $z$ = 1), and the code should include the formulation of the load model $P$ and $Q$ considering constant impedance, as follows \cite{667386}:
\begin{subequations} 
	\label{keyE6.21211} 
	\begin{align}
		P=&S_0\cos(\theta)(\frac{V}{V_{lim}})^2 \label{eq:line1}\\
		Q=&S_0\sin(\theta)(\frac{V}{V_{lim}})^2 \label{eq:line2}
	\end{align}
\end{subequations}
where $V$ is the magnitude of the voltage on the load, $S_0$ is the load apparent power with voltage $V_{lim}$, using $V_{lim}$ equal to $V_{max}$ or $V_{min}$, and $\cos(\theta)$ the load power factor. The variable $u = 1 $ means that the load is connected to the system.
\begin{table}[h]
	\renewcommand{\arraystretch}{1.1}
	\centering
	\caption{Data structure of PQ buses (PQ.con)}
	\label{tabela 3}
	\begin{tabular}{|c|c|l|c|}
		\hline 
		\textit{Column} & \textit{Variable} & \textit{Description} & \textit{Unit} \\
		\hline 
		1 & $\#$ & \textit{Bus number} & \textit{int} \\ 
		
		2 & $S_b$ & \textit{Power base} & \textit{MVA} \\ 
		
		3 & $V_b$ & \textit{Voltage base} & \textit{kV}  \\ 
		
		4 & $P_{L}$ & \textit{Load active power}  & \textit{p.u.} \\ 
		
		5 &  $Q_{L}$& \textit{Load reactive power} & \textit{p.u.} \\ 
		
		6 &  $V_{max}$& \textit{Maximum voltage} & \textit{p.u.} \\ 
		
		7 &  $V_{min}$& \textit{Minimum voltage} & \textit{p.u.} \\  
		
		8 & z & \textit{Allow conversion to impedance }& \textit{boolean} \\
		
		9 & u & \textit{Connection status}& \textit{boolean} \\
		\hline 
	\end{tabular} 
\end{table}
\section{Models of wind, hydro, fossil fuel power plants, reactive power compensation devices, and their data structures}

This section presents the capacity factors (CFs) of wind, hydro, and fossil fuel  power plants, as well as the models of the buses where these power plants are connected. In addition, it describes the data structure of the voltage control buses (PV) used for the synchronous generating units of hydro and fossil fuel power plants, and the PQ generators buses (PQgen) used for the wind turbines, mostly asynchronous, present in the wind farms. The data structure of the slack bus (SW) is also described.	
\subsection{Modeling of the wind farms capacity factor} 
This work uses the realistic and updated CFs, specially considering the current drought faced in the last years.  As a result, this test system employs average monthly capacity factors, by primary energy source, and by subsystem, obtained from historical data of centralized power plants in each subsystem of the BIPS, which are provided by the EPE in \cite{ASH2:2021:Online}.	
\par The application of CFs, monthly and by primary source, allows to simulate the most accurate generation for all power plants due to the influence of the seasonality of each source. Thus, it is possible to carry out reliable static analyzes with the NE-BIPS-229bus test system, and compatible with the different load levels of the NE subsystem of the BIPS, which were also discretized by month and by subsystem, as described in the section  \ref{sec:modelagem-dos-patamares-de-cargas-do-sin}.	
\subsection{Capacity factors data structure}
In order to have a uniform data structure for all primary energy sources, the 
Table~\ref{tabela 23} presents the structure for capacity factors which are common for wind  $(\textit{CFW}_m)$, hydro $(\textit{CFH}_m)$ and fossil fuel  power plants $(\textit{CFF}_m)$. 
\begin{table}[h]
	\renewcommand{\arraystretch}{1.1}
	\centering
	\caption{Capacity factors data structure of wind/hydro/fossil fuel  power plants (CapacFactor.Wind/Hydro/FFuel.con)}
	\label{tabela 23}
	\begin{tabular}{|c|c|l|c|}
		\hline
		\textit{Column} & \textit{Variable} & \textit{Description} & \textit{Unit} \\ 
		\hline 
		1 & $A_i$ & \textit{Area number} & \textit{int} \\ 
		2 & \textit{${CFW}_m$}/\textit{${CFH}_m$}/\textit{${CFF}_m$} & Capacity factor & \textit{real} \\ 
		\hline
	\end{tabular}
\end{table}
\begin{table*}[h]
	\renewcommand{\arraystretch}{1.1}
	\centering
	\caption{Data of Capacity Factor of Wind Power Plants (CapacFactor.Wind.con)}
	\label{tabela 24}
	\begin{tabular}{|c|c|c|c|c|c|c|c|c|c|c|c|c|c|}
		\hline 
		\textit{Area number}&\textit{${CFW}_1$} &\textit{${CFW}_2$}& \textit{${CFW}_3$}& \textit{${CFW}_4$}& \textit{${CFW}_5$} &\textit{${CFW}_6$}& \textit{${CFW}_7$}& \textit{${CFW}_8$}&\textit{${CFW}_9$}&\textit{${CFW}_{10}$}& \textit{${CFW}_{11}$}& \textit{${CFW}_{12}$}\\
		\hline
		3& 0.39& 0.39& 0.32& 0.37&	0.45 &0.52 &0.56& 0.6 &0.61&	0.54& 0.49 &0.42\\
		\hline
	\end{tabular}
\end{table*}
The subsystem number is structured  in the first row of Table~\ref{tabela 23}, which is referred as area in the test system. In the second row, the capacity factors by source for each month $m$ are structured. For example, Table~\ref{tabela 24} presents the data of capacity factors of the wind farms of the NE subsystem (\#\hspace{0.05cm}Area 3) which presents the highest capacity factors of the BIPS, and they are even higher in the period from June to November. All data about the capacity factors of hydro and fossil fuel  power plants has the same structure of wind farms and are in the NE-BIPS-229bus. Thus, the formulation which can be used to adjust the generation of a specific power plant, such as a wind farm, is a function of its capacity factor for a specific month $m$ as follows:
\begin{subequations} 
	\label{keyE3} 
	\begin{align} 			
		P_{gm} &= \textit{CFW}_mP_{g} \label{eq:line31}\\
		Q_{gm} &= \textit{CFW}_mQ_{g} \label{eq:line32}		
	\end{align}
\end{subequations}
\subsection{Modeling of wind farms}\label{sec:the-wind-farm-generation-bus-modeling}
Variable speed wind turbines, as the doubly-fed induction generators (DFIG) and the full converter wind generators are currently the most used technology in wind farms. They are composed of the Rotor-side converter (RSC) and the Grid-side converter (GSC), which may operate temporarily overloaded to contribute to the system's reactive power support during short-term disturbances. 
However, for steady-state operation, DFIG equipped with Insulated Gate Bipolar Transistors (IGBT), that becomes increasingly popular in power control circuits for industrial use, have at synchronous speed operating point a limitation arised due to the maximum junction temperature of the IGBTs. This effect causes a reduction in the maximum allowed output current in the RSC \cite{5994933}. Consequently, it limits the contribution in reactive power generation, that also limits the allowable power factor (PF) range of the generator. 
\par In Brazil, for example, the Brazilian Electricity Regulatory Agency (ANEEL) determines that the PF of wind farms should be in the range of 0.95 inductive and 0.95 capacitive. Thus, this PF range determines the contribution limit of each DFIG wind turbine in the reactive power support at the point of interconnection of the system where a wind farm is coupled, aiming at the static voltage stability assessment.

The terminal voltage of the wind farms, which affects the capability curve, are set in the range of 0.9\hspace{0.1cm}-\hspace{0.1cm}1.1 p.u., as proposed in \cite{7232809}.
The terminal buses of wind farms are modeled as PQ buses considering a negative load, which are one of the simplest representation of wind farms for steady-state analyses \cite{5159366}. On the other side, wind farms may also be modeled as a voltage controlled bus (PV) including the operational limits, that is, according to the capability curve. In this paper, the PQ bus model with negative load was employed.  
\par As a result, in the data structure of the test system NE-BIPS-229bus, the wind farms are modeled as PQ buses with a negative load, with voltage limits of 0.9 - 1.1 pu, and power factor in the range of 0.95 inductive and 0.95 capacitive.

\subsection{Data structure for PQ buses as wind farms}
As described in section \ref{sec:the-wind-farm-generation-bus-modeling}, the PQ generator buses are modeled as constant active and reactive power, as long as the voltages are in the limits. If one of the voltage limits is violated, the PQ generator buses can be converted to constant impedance $z$ according to the equations (\ref{eq:line1}) and (\ref{eq:line2}), but only during the iterative process of the power flow (LF), continuation
power flow (CPF), and optimal power flow (OPF) \cite{milano2006open}. 
It is important to highlight that, as the PQ generators are seen by the routines as negative loads, it is not allowed to connect PQ loads to the same bus of a PQ generator.	
\par The data structure of PQ generators (PQgen.con), detailed in Table~\ref{tabela 6}, is the same data structure of PQ load buses, the difference is that the loads $P_L$ and $Q_L$ are replaced by the fixed active power $P_g$ and reactive power $Q_g$. In this work, the active and reactive powers have positive signal in the structure PQgen.con, and as these values are constant, the ratio between $P_g$ and $Q_g$ has to comply with the required power factor range, which is between 0.95 inductive to 0.95 capacitive, considering a voltage limit in the range of 0.9 and 1.1 p.u.

\begin{table}[h]
	\renewcommand{\arraystretch}{1.1}
	\centering
	\caption{PQ generator data structure  (PQgen.con)}
	\label{tabela 6}
	\begin{tabular}{|c|c|l|c|}
		\hline 
		\textit{Column} & \textit{Variable} & \textit{Description} & \textit{Unit} \\
		\hline 
		1 & $\#$ & \textit{Bus number} & int \\ 
		
		2 & $S_b$ & \textit{Power base} & \textit{MVA} \\ 
		
		3 & $V_b$ & \textit{Voltage base} &\textit{kV}  \\ 
		
		4 & $P_{g}$ & \textit{Active power}  & \textit{p.u.} \\ 
		
		5 &  $Q_{g}$& \textit{Reactive power} & \textit{p.u.} \\ 
		
		6 &  $V_{max}$& \textit{Maximum voltage} & \textit{p.u.} \\ 
		
		7 &  $V_{min}$& \textit{Minimum voltage} & \textit{p.u.} \\  
		
		8 & z & \textit{Allow conversion to impedance}& \textit{boolean} \\
		
		9 & u & \textit{Connection status}& \textit{boolean} \\
		\hline 
	\end{tabular} 
\end{table}	
Detailed and updated data of wind farms are available online in the data base Sigel from ANEEL in \cite{ASH:20216:Online}. As an example, Table~\ref{tabela 7} which presents part of the data from the PQ generator buses with their numerical generation values to meet the annual average load base case of the NE subsystem, and which appear in the PQgen.con structure of the NE-BIPS-229bus test system, includes 101-CQBRD1EOL034 (Canoa Quebrada wind farm). The active power is 0.3477 p.u. (0.57x$CFW_9(3)$) and the reactive is null, that is, for the power flow (without optimization) the unit power factor is fixed for all wind farms. When the simulations for OPF, presented in the section   \ref{sec:modelos-de-ofertas-e-custos-de-geracao-e-demanda-e-suas-estruturas-de-dados}, it can be observed that the resolution itself will seek the generation of optimal reactive power. 

\begin{table}[h]
	\renewcommand{\arraystretch}{1.1}
	\centering
	\caption{Data from PQ generator (PQgen.con)}
	\label{tabela 7}
	\begin{tabular}{|c|c|c|c|c|c|c|c|c|} 
		\hline 
		$\#$ & $S_b$  & $V_b$  &   $P_{g}$   & $Q_{g}$     &  $V_{max}$ &  $V_{min}$  &   z   & u \\
		\hline
		101     & 100   & 34.5  & 0.3477     & 0     & 1.1   & 0.9   & 0     & 1 \\
		\hline 
	\end{tabular} 
\end{table}
\subsection{Data structure of the reference slack bus}\label{sec:estrutura-de-dados-da-barra-de-referencia}	
The bus that represents the generators that are reference are modeled as fixed terminal voltage magnitude and angle, as shown in Table~\ref{tabela 8}.	
\begin{table}[h]
	\renewcommand{\arraystretch}{1.1}
	\centering
	\caption{Data structure of slack bus (SW.con)}
	\label{tabela 8}
	\begin{tabular}{|c|c|l|c|}
		\hline 
		\textit{Column} & \textit{Variable} & \textit{Description} & \textit{Unit} \\
		\hline 
		1 & $\#$ & \textit{Bus number} & \textit{int} \\ 
		
		2 & $S_b$ & \textit{Power base} & \textit{MVA} \\ 
		
		3 & $V_b$ & \textit{Voltage base} & \textit{kV}  \\ 
		
		4 &  $V_0$ & \textit{Voltage magnitude} & \textit{p.u.} \\ 
		
		5 & $\theta_0$ & \textit{Phase reference} & \textit{rad} \\ 
		
		6 & $Q_{max}$& \textit{Maximum reactive power} & \textit{p.u.} \\ 
		
		7 & $Q_{min}$& \textit{Minimum reactive power} & \textit{p.u.} \\ 
		
		8 &  $V_{max}$& \textit{Maximum voltage} & \textit{p.u.} \\ 
		
		9 &  $V_{min}$& \textit{Minimum voltage} & \textit{p.u.} \\  
		
		10 &$P_{g0}$ & \textit{Initial active power} & \textit{p.u.} \\ 
		
		11 &$\gamma$ & \textit{Loss-sharing factor} & \textit{[0 1]} \\ 
		
		12 & z & \textit{Reference bus}& \textit{boolean} \\ 
		
		13 & u & \textit{Connection status}& \textit{boolean} \\
		\hline 
	\end{tabular} 
\end{table}
\par Table \ref{tabela 8} presents the $P_{g0}$ and $\gamma$ for the reference bus, which the user should use in CPF and OPF, when using the distributed reference bus. $P_{g0}$ represents the initial active power (for CPF e OPF) generated and that it is not included in the short-term marked hired in the Free Market of Energy (in the case of Brazil) or from market bidding \cite{milano2005voltage} to supply the demand. Thus, the generated power should respect:	
\begin{align}
	P =(1 +\gamma k_G)P_{g0}  \label{eq:line3}
\end{align}
\noindent where $\gamma$ is the loss-sharing factor, and $k_G$ represents a scale used to distribute the system losses, associated to the solution of power flow equations \cite{1198291}.
In the case $\gamma$ is not specified, it is assumed as $\gamma$ = 1. On the other side, if more than 1 bus is set as slack bus, only one will have a fixed phase and, in this case, $z$ from SW.con should be 1 and for the other generators $z$ = 0. In the proposed test system, the hydro power plant of Xingó (\# bus 294), which is one of the largest power plants of the studied subsystem, is adopted as the reference. This power plant presents a capacity of 3.162 GW, with a capacity factor, forecasted for September of 2021, of 32\% ($CFH_9(3)$).	

\subsection{PV buses data structure for hydro and fossil fuel plants}	
Table~\ref{tabela 5} shows the required data for the structure of PV buses. As one can see, typical data as bus number, power and voltage base, as well as voltage magnitude and active power injected at the bus. For CPF and OPF, voltage and reactive power limits are also in the structure. In the case the code implemented by the user employs the distributed reference bus model, the variable $\gamma$, the loss-sharing factor, should be subjected to $ 0 < \gamma < 1$, and the generated power $P$ should satisfy the following expression:
\begin{align}
	P =(1 +\gamma k_G)P_{g}  \label{eq:line4}
\end{align}
For PV buses, $\gamma=0$ if the adopted model is not the distributed slack bus.	
\begin{table}[h]
	\renewcommand{\arraystretch}{1.1}
	\centering
	\caption{Data structure for PV buses (PV.con)}
	\label{tabela 5}
	\begin{tabular}{|c|c|l|c|}
		\hline 
		\textit{Column} & \textit{Variable} & \textit{Description} & \textit{Unit} \\
		\hline 
		1 & $\#$ & \textit{Bus number} & \textit{int} \\	
		2 & $S_b$ & \textit{Power base} & \textit{MVA} \\	
		3 & $V_b$ & \textit{Voltage base} &\textit{kV}  \\ 	
		4 & $P_{g}$ & \textit{Active power}  & \textit{p.u.} \\ 	
		5 &   $V_0$ & \textit{Voltage magnitude} & \textit{p.u.}  \\ 	
		6 & $Q_{max}$& \textit{Maximum reactive power} & \textit{p.u.} \\	
		7 & $Q_{min}$& \textit{Minimum reactive power} & \textit{p.u.} \\ 	
		8 &  $V_{max}$& \textit{Maximum voltage} & \textit{p.u.} \\ 	
		9 &  $V_{min}$& \textit{Minimum voltage} & \textit{p.u.} \\  	
		10 &$\gamma$ & \textit{Loss-sharing factor} & \textit{[0 1]} \\ 	
		11 & u & \textit{Connection status}& \textit{boolean} \\
		\hline 
	\end{tabular} 
\end{table}

\subsection{Shunt power compensation data structure}
The shunt active and reactive power compensation devices are modelled by adding its admitance, with respective conductance $g$ and susceptance $b$ to the admitance matrix $Y$. The relationship of $g$ and $b$ with the active and reactive powers injected by the compensators in the bus where they are coupled are given by: 
\begin{subequations} 
	\label{keyE6.31211} 
	\begin{align}
		P =-gV^2 \label{eq:line5}\\
		Q =-bV^2 \label{eq:line6}
	\end{align}
\end{subequations} 
\noindent where $V$ is the bus voltage. The susceptance $b$ is negative for inductive loads, and positive for capacitive. The data structure for the shunt compensation devices (Shunts.con) is presented in Table~\ref{tabela 14}, where the variables are detailed. Shunt allocation is a very useful feature in wind farms operation.	
\begin{table}[h]
	\renewcommand{\arraystretch}{1.1}
	\centering
	\caption{Data structure of shunt compensation devices (Shunts.con)}
	\label{tabela 14}
	\begin{tabular}{|c|c|l|c|}
		\hline 
		\textit{Column} & \textit{Variable} & \textit{Description} & \textit{Unit} \\ 
		\hline  		
		1 & \# & \textit{Bus number} & \textit{int} \\ 
		
		2 & $S_b$ & \textit{Power base} & \textit{MVA} \\ 
		
		3 & $V_b$ & \textit{Voltage base} &\textit{kV}  \\ 
		
		4 & $f_n$ & \textit{Nominal frequency} & \textit{Hz} \\ 
		
		5 & g & \textit{Condutance} & \textit{p.u.}\\ 
		
		6 & b & \textit{Susceptance} & \textit{p.u.} \\ 
		
		7 & u & \textit{Connection status}& \textit{boolean} \\
		\hline 
	\end{tabular}
\end{table}
\par As an example, Table~\ref{tabela 15} shows the numerical data of a compensator (fictitious, for testing) coupled to buses 104 (ARACT2-CE034) and 112 (RUSSAS-CE069), identified as critical in simulations performed in the NE-BISP-229bus test system. This research did not obtain accurate data from the compensators that exist in the NE subsystem, so it is considered as fictitious ($b$ = 0.1), and in the test system it has the status of disconnected ($u$ = 0), but the user can connect it to any bus for simulations.
  \begin{table}[h]
	 	\renewcommand{\arraystretch}{1.1}
	 	\centering
	\caption{Compensators data (Shunts.con)}
		\label{tabela 15}
	\begin{tabular}{|c|c|c|c|c|c|c|}
	\hline 
	 $\#$ & $S_b$  & $V_b$ & $f_n$  &   g &  b & u \\
	\hline 
	103    & 100 & 34.5 &   60    & 0  & 0.1 & 0 \\
		104    & 100 & 34.5 &   60    & 0  & 0.1 & 0 \\
	\hline 
	\end{tabular}%
	\end{table} 
	\section{Modeling of transmission and subtransmission lines, transformers, and data structure}	
	This section presents the model adopted for the transmission and transformers systems, and describes its data structure and compiled data. The transmission (500 e 230 kV) and subtransmission (69, 34.5, 20 e 13.8 kV) lines and transformers are modeled as lumped parameters with the equivalent $\pi$ circuit  \cite{monticelli1983fluxo}. Table \ref{tabela 16} shows the data structure for transmission and subtransmission lines, and transformers (Line.con), using either the parameters in units per km or in p.u. When p.u. is used, $\textit{l}$ (line length) is null. When $\textit{l} > 0$, resistance $r$, reactance $x$, and susceptance $b$ are expressed in units per km. The lumped parameter corresponds to the line susceptance $b$ divided by 2, allocated at each end of the line, either in units per km or in p.u.
	\par It is assumed that when the variable $k_{T}$ of Table \ref{tabela 16}, which is used as data of the transformation ratio between the primary and secondary of a transformer $kV/kV$, is fixed as $k_ {T}=0$, means that this data line of the data structure represents a transmission or subtransmission line. If $k_{T}>0$, the data line represents a transformer. The same criterion is applied, together, for the variable $a$ that indicates the transformer tap ratio, that is, if $a = 1$ it is a transformer with nominal tap, and if $a = 0$, it is a transmission line. The variable $\phi$ represents the phase of the phase-shifting transformer. The variables $I^{max}$, $P^{max}$, and $S^{max}$ are respectively the current, active power, and apparent power limits of line or transformer.	
	\begin{table}[h]
		\renewcommand{\arraystretch}{1.1}
		\centering
		\caption{Data structure for transmission line and transformers (Line.con)}
		\label{tabela 16}
		\begin{tabular}{|c|c|l|c|}
			\hline 
			\textit{Column} & \textit{Variable} & \textit{Description} & \textit{Unit} \\ 
			\hline 
			1 & i & \textit{From Bus} & \textit{int} \\ 
			
			2 & j & \textit{To Bus} & \textit{int} \\ 
			
			3 & $S_b$ & \textit{Power base } & \textit{MVA} \\ 
			
			4 & $V_b$ & \textit{Voltage base} &\textit{kV}  \\ 
			
			5 & $f_n$ & \textit{Nominal frequency} & \textit{Hz} \\ 
			
			6 & \textit{l} & \textit{Line length} & \textit{km} \\ 
			
			7 & $k_{T}$ & \textit{Transformation ratio P/S} & \textit{kV/kV} \\ 
			
			8 & r & \textit{Resistance} & \textit{p.u. (}$\Omega$\textit{/km)} \\ 
			
			9 & x & \textit{Reactance} & \textit{p.u. (H/km)} \\ 
			
			10 & b & \textit{Susceptance} & \textit{p.u. (F/km)} \\ 
			
			11 & a & \textit{Ratio of the transformer tap} & \textit{p.u./p.u.}\\ 
			
			12 & $\phi$ & \textit{Transformer phase shift}&  \textit{°(deg)}\\ 
			
			13 & $I^{max}$ & \textit{Current limit}&\textit{p.u.} \\ 
			
			14 & $P^{max}$  & \textit{Active power limit} & \textit{p.u.} \\ 
			
			15 & $S^{max}$ & \textit{Apparent power limit} & \textit{p.u.} \\ 
			
			16 & u & \textit{Connection status} & \textit{boolean} \\ 
			\hline 
		\end{tabular} 
	\end{table}	
	\par For example, Table ~\ref{tabela 17} shows some data of line and transformer (Line.con). It can be observed  that there is a 230 kV transmission line from bus 105 to 106, since columns 7 and 11 are null, and a transformer  between 104 and 105.
	\begin{table*}[h]
		\renewcommand{\arraystretch}{1.1}
		\centering
		\caption{Data from transmission lines and transformers (Line.com)}
		\label{tabela 17}
		\begin{tabular}{|c|c|c|c|c|c|c|c|c|c|c|c|c|c|c|c|}
			\hline 
			k & m & $S_b$ &  $V_b$ & $f_n$ &  \textit{l} & $k_{T}$& r & x & b & a & $\phi$ &$I_{max}$ &$P_{max}$ &$S_{max}$ & u\\
			\hline 
			104& 105& 100& 34.5& 60& 0& 34.5/230& 0& 0.125& 0& 1/1& 0& 0.46& 0.72& 0.8& 1\\
			105& 106& 100& 230& 60& 0& 0& 0.0076& 0.0576& 0.052& 0& 0& 1.16& 1.8& 2& 1\\ 		
			\hline 
		\end{tabular}%
	\end{table*}%
	\section{Supply costs and demand prices models, and their data structures }\label{sec:modelos-de-ofertas-e-custos-de-geracao-e-demanda-e-suas-estruturas-de-dados}
	This section addresses the modeling of supply parameters, generation costs and prices applied to demand that are not part of the Regulated Contracting Environment, known as Ambiente de Contratação Regulada - ACR, and that must be contracted in the Free Energy Market (in the case of Brazil), or market bidding \cite{milano2005voltage}, to meet the additional demand for power and energy. To facilitate the description of these parameters, the formulation of the multiobjective optimization of market and voltage security \cite{1198291} is described in this section. 	
	\par The generation cost and demand price parameters, in addition to market constraints, which are included in the supply (Supply.con) and demand (Demand.con) data structures \cite{milano2006open}, described in this section, can be used specifically in OPF routines with voltage security constraints (VSC) that serve to model prices in the Free Energy Market. Besides, it can also be used for voltage stability analysis, or both analysis when applied to multi-objective optimization \cite{1198291}. These supply and demand data structures predict the input of parameters that determine the generation and load growing direction, which are useful in CPF simulations, among others. 
	\par This test system adopts the assumption proposed in \cite{1198291}, which proposes that the inelastic demand, that is, when the load cost do not vary with load increase, this information is included in the load bus data structure (PQ.con). It is also assumed that to meet these loads (including losses) the hydro and fossil fuel power plats are part of the data structure of slack bus (SW.con) and voltage control buses (PV.con), and the wind farms are part of the PQgen.con structure.	
	\par This assumption allows fossil fuel plants, due to their high cost, to only be dispatched (if necessary) to the short-term market, and as a result of the OPF to be applied.  Additionally, another assumption adopted from \cite{1198291} is that the inelastic loads corresponds to 90\% of the load of the base case, resulting in 10\% of the load supplied by the short-term market. To make such assumptions viable, the supply data structure (Supply.con) is used to include all the power plants that participate in the short-term market, or market bibding, and the demand data structure (Demand.con) used for elastic loads.	
	\subsection{Supply data structure}	
	The supply data structure (Supply.con), detailed in Table \ref{tabela 18}, which contains the data for offers and costs of active and reactive generation, are used with demand data structure (Demand.con), Table \ref{tabela 19}, in CPF and OPF scripts, among others. 
	\begin{table}[h]
		\renewcommand{\arraystretch}{1.1}
		\centering
		\caption{Supply data structure (Supply.con)}
		\label{tabela 18}
		\begin{tabular}{|c|c|l|c|}
			\hline 
			\textit{Column} & \textit{Variable} & \textit{Description} & \textit{Unit} \\
			\hline 
			1 & \# & \textit{Bus number} & \textit{int} \\ 
			
			2 & $S_b$ & \textit{Power base} & \textit{MVA} \\ 
			
			3 & $P_{S_0}$ & \textit{Active power growth direction} & \textit{p.u.} \\  
			
			4 & $P_{S}^{max}$ & \textit{Maximum offer of active power} & \textit{p.u.} \\ 
			
			5 & $P_{S}^{min}$ & \textit{Minimum offer of active power} & \textit{p.u.} \\ 
			
			6 & $P_{S}$ & \textit{Optimized offered active power} & \textit{p.u.} \\ 
			
			7 & $C_{P_0}$ & \textit{Fixed cost (active power)} & \textit{R\$/h}  \\ 
			
			8 & $C_{P_1}$ & \textit{Linear cost (active power)} &\textit{R\$/MWh}   \\ 
			
			9 & $C_{P_2}$  & \textit{Quadratic cost (active power)} &  \textit{${R\$/MW}^2$}\textit{h} \\ 
			
			10 & $C_{Q_0}$ & \textit{Fixed cost (reactive power)} & \textit{R\$/h}  \\ 
			
			11 & $C_{Q_1}$ & \textit{Linear cost (reactive power)} & \textit{R\$/MVArh }  \\ 
			
			12 & $C_{Q_2}$ &\textit{Quadratic cost (reactive power)} & \textit{${R\$/MVAr}^2$}\textit{h} \\ 
			
			13 & $u_c$ & \textit{Unit commitment variable} & \textit{boolean} \\ 
			
			14 & Fut. use&  & \\ 
			
			15 & $\gamma$ & \textit{Loss-sharing factor} & - \\ 
			
			16 & $Q_g^{max}$& \textit{Maximum reactive power} & \textit{p.u.} \\ 
			
			17 & $Q_g^{min}$& \textit{Minimum reactive power} & \textit{p.u.} \\ 
			
			18-19 & Fut. use &  & \\ 
			
			20 & u & \textit{Connection status}& \textit{boolean} \\
			\hline 
		\end{tabular} 
	\end{table}	
	The multi-objective optimization function $G$ (\ref{eq:line7}), that uses the data from these structures,  presents two terms, which have influence on the final solution and depend on the weighting factor $\omega$. The first term searches for the  short-term market solution, and the second term, $(1-\omega)\lambda_c$, searches for the maximization of the voltage stability margin. Usually, $0 \le \omega \le  1$, but values of $\omega$ very close to unity have been adopted to minimize the influence of the term $(1-\omega)\lambda_c$ on the market optimization \cite{1198291}, and values of $\omega$ as 0 or 1 so that the optimization is restrained to one of the terms of $G$. Besides the constrains presented in (\ref{keyE6.111}), other  related to the operating limits of the system, not presented in this formulation, are detailed in \cite{1198291}.
	\begin{align}
		Min.  ~  G =  -\omega(C_D(P_{D}) - C_S(P_S))-(1-\omega)\lambda_c \label{eq:line7}
	\end{align}
	$\hspace{2cm} s.t.$
	\begin{subequations} 
		\label{keyE6.111} 
		\begin{align}		
			f_i(x,Q_G,P_{G},P_{L}) = 0,\hspace{0.cm} \forall i \in \hspace{0.0cm} \mathcal{I} \\
			\label{keyE6.121}
			f_i(x_c,Q_{G_c},\lambda_c,P_{S_s},P_{D_e}) = 0,\hspace{0.0cm} \forall i \in \hspace{0.0cm} \mathcal{I}\\
			\label{keyE6.131}
			\lambda_{min} \le \lambda_c \le \lambda_{max} \\
			\label{keyE6.141}
			P_{S_{s}}^{min} \le P_{S_s} \le P_{S_{s}}^{max},\hspace{0.0cm}\forall s \in \hspace{0.0cm} \mathcal{G}_s \hspace{0.0cm}\\
			\label{keyE6.1411}
			P_{D_e}^{min} \le P_{D_e} \le P_{D_e}^{max},\hspace{0.0cm}\forall e \in \hspace{0.0cm} \mathcal{D}_{elas} \hspace{0.0cm}
		\end{align}
	\end{subequations}
	\noindent where:
	\begin{itemize}	
		\item $C_S(P_{S})$ represents the total cost per hour of generation to supply the short-term market, which can include, besides the corresponding active power generation, the cost of reactive power (which can be used as ancillary service). $C_S(P_{S})$ is a quadratic function, which include the variable $P_S$ (to be optimized by the OPF) and the cost coefficients $C_{P_0}$ to $C_{Q_2}$ from Table~\ref{tabela 18}, according to the formulation presented in the PSAT manual \cite{milano2006open}. 
		\item $C_D(P_{D})$ represents the demand total price per hour in the short-term market, which can include, besides the pricing of the active power, the reactive power pricing. $C_D(P_{D})$ is a quadratic function of the variable $P_D$, which represents the demand vector of elastic loads (to be optimized by the OPF), and from the cost coefficients $C_{P_0}$ to $C_{Q_2}$ of Table~\ref{tabela 19}.
		\item $\mathcal{I} = \{1, 2, ..., N_B\}$, is the set of system buses, and $N_B$ is the total number of buses. 
		\item $\mathcal{G}_s = \{s_1, s_2, ..., {s_p}\}$, 
		is the set of power plants selected by some criteria to be available in the Supply.con structure, and $p$ is the total number of power plants selected.
		\item $f_i(x,Q_G,P_{G},P_{L})\hspace{0.05cm}=\hspace{0.05cm}0$, are the power balance equations for the base case charge, for the $i$-th bus.
		\item $f_i(x_c,Q_{G_c},\lambda_c,P_{S_s},P_{D_e})\hspace{0.01cm}=\hspace{0.01cm}0$, is the second set of power flow equations, with subscript $c$ introduced to represent the system in critical conditions associated to the maximum loadability, where $\lambda$ is the parameter of load increase.
		\item The usual value for $\lambda_{min}$ is 1.01 to ensure a minimum voltage stability margin, and a maximum of 1.99 for $\lambda_{max}$ to avoid that an excessive margin requirement compromise the prices optimization. $\lambda$ = 1 represents the base case loading.
		\item It is necessary to present to the user that wants to implement the OPF rotines, the following formulations to active power generation and demand, not only for the assumption of 90\% of the base case ($P_{G_0}$ e $P_{L_0}$), but also for the base case ($P_G$ e $P_L$), and for critical conditions ($P_{G_c}$ and $P_{L_c}$) of the point of maximum loadability (PML), as follows \cite{1198291}:
		\begin{subequations} 
			\label{keyE6.112} 
			\begin{align} 			
				P_G &= P_{G_0} + P_{S_s} \label{eq:line8}\\
				P_L &= P_{L_0} + P_{D_e} \label{eq:line10}\\
				~P_{G_c}& =(\lambda_c + k_{G_c})P_G \label{eq:line11}\\
				P_{L_c}& = \lambda_c P_L  \label{eq:line12}			
			\end{align}
		\end{subequations}
		\noindent where, $P_{G_0}$ e $P_{L_0}$ represent generation and load, respectively, which are not included in the market bidding, and $k_{G_c}$ is a scalar variable to share the system losses when operating in critical conditions of PML.	
		\item However, if the user implement the CPF routines to analyze bifurcations, the following formulation of active power generation and active and reactive power loads should be considered in place of (\ref{eq:line8})-(\ref{eq:line10}):	
		\begin{subequations} 
			\label{keyE6.1121} 
			\begin{align} 			
				P_G &= P_{G_0} + (\lambda -1 +\gamma k_G)P_{S_0} \label{eq:line81}\\
				P_L &= P_{L_0} + (\lambda -1) P_{D_0} \label{eq:line101}\\
				Q_L &= Q_{L_0} + (\lambda -1) Q_{D_0}		
			\end{align}
		\end{subequations}
       where $P_{G_0}$, $P_{L_0}$, and $Q_{L_0}$ are the active power generation, and the active and reactive power load, respectively, of base case. $P_{S_0}$ is active power growth direction, $\gamma$ is the loss-sharing factor, which are part of the data structure Supply.con, Table~\ref{tabela 18}, and $k_G$ presented in section \ref{sec:estrutura-de-dados-da-barra-de-referencia}. On the other side, $P_{D_0}$ and $Q_{D_0}$ are the directions of active and reactive load growth, which are in the data structure  Demand.con, Table~\ref{tabela 19}.  
		\item $x = [\theta,V]$ and $x_c = [\theta_c,V_c]$ are the vectors with system state variables, \textit{i.e.}, the angles and magnitudes of the voltages, for the base case and the PML, respectively.
		\item $Q_G$ and $Q_{G_c}$ are the vectors with the power plants reactive power generations, for the base case and PML, respectively.
		\item $P_{S_s}$, active power generation from the $s$ power plant selected by some criteria to be available in the Supply.con structure to participate in the market bidding, which will be optimized by the OPF in the range of $P_{S}^{min}$ and $P_{S}^{max}$ available in the structure Supply.con in Table~\ref{tabela 18}. This criteria could use the ranking of the sensitivities index of wind farms that must be adjusted their power factor, as voltage preventive control actions aiming to contribute in the loading margin improvement, as detailed in \cite{{9638214,silva2019loading}}. 
		\item  $P_{D_e}$ represents the demand vector of the set $\mathcal{D}_{elas}$  of $e$ elastic loads, which will be optimized in the range $P_{D}^{max}$ and $P_{D}^{min}$ (Table~\ref{tabela 19}), and that corresponds to the PQ buses with elastic loads included in the structure Demand.con detailed in the next section. 
		\end{itemize}
		\par The test system uses  $C_{P_1}$, that is, as a linear cost coefficient of active power generation, the Unitary Variable Cost, known as Custo Variável Unitário - CVU, of one of the combined cycle power plants (gas and steam) operating in NE subsystem, the Termopernambuco  (SUAPE), which currently (July 2021) presents a cost of 173.38\hspace{0.05cm}R\$/MWh (about 33.00\hspace{0.05cm}\$/MWh), which is used by the Electric Energy Commerce Chamber (CCEE) \cite{ASH4:2021:Online} and regulated by ANEEL to verify the generation. This value is included in the data structure Supply.con as $C_{P_1}$. The other cost coefficients are null since it was not possible to access to this information, but the data structure allows including these values. 
		\subsection{Demand data structure}
		The demand data structure (Demand.con) comprehends demand data considering cost coefficients applied to short-term market. It should be provided the maximum $P_{D}^{max}$ and minimum demand bid $P_{D}^{min}$, as well as the costs coefficients $ C_{P_0}$ to $C_{Q_2}$, which are for both active and reactive power, used by the OPF and CPF routines. The load growth direction of active $P_{D_0}$ and reactive power $Q_{D_0}$ should also be included, which are also used in the OPF and CPF routines. If $P_{D_0}$ and $Q_{D_0}$ are not available, active and reactive power of the load from base case should be used. The data structure is detailed in Table \ref{tabela 19}.
			\begin{table}[h]
			\renewcommand{\arraystretch}{1.1}
			\centering
			\caption{Demand data structure (Demand.con)}
			\label{tabela 19}
			\begin{tabular}{|c|c|l|c|}
				\hline 
				\textit{Column} & \textit{Variable} & \textit{Description} & \textit{Unit} \\
				\hline 
				1 & \# & \textit{Bus number} & \textit{int} \\ 
				
				2 & $S_b$ & \textit{Power base } & \textit{MVA} \\ 
				
				3 & $P_{D_0}$ & \textit{Active load growth direction} & \textit{p.u.} \\ 
				
				4 & $Q_{D_0}$ & \textit{Reactive load growth direction} & \textit{p.u.} \\ 
				
				5 & $P_{D}^{max}$ & \textit{Maximum demand of active power} & \textit{p.u.} \\ 
				
				6 & $P_{D}^{min}$ & \textit{Minimum demand of active power} & \textit{p.u.} \\ 
				
				7 & $P_{D}$ & \textit{Optimized active power demand} & \textit{p.u.} \\ 
				
				8 & $C_{P_0}$ & \textit{Fixed cost (active power)} & \textit{R\$/h} \\
				
				9 & $C_{P_1}$ & \textit{Linear cost (active power)} & \textit{R\$/MWh} \\
				
				10 & $C_{P_2}$ & \textit{Quadratic cost (active power)} & \textit{${R\$/MW}^2$}\textit{h}\\
				
				11 & $C_{Q_0}$ & \textit{Fixed cost (reactive power)} & \textit{R\$/h} \\
				
				12 & $C_{Q_1}$ & \textit{Linear cost (reactive power)} & \textit{R\$/MVArh}\\
				
				13 & $C_{Q_2}$ & \textit{Quadratic cost (reactive power)} & \textit{${R\$/MVAr}^2$}\textit{h} \\
				14 & $u_c$ & \textit{Unit commitment variable} & \textit{boolean} \\ 
				
				15-17 & Fut. use &  & \\ 
				
				
				
				18 & u & \textit{Connection status}& \textit{boolean}\\
				\hline 
			\end{tabular} 
		\end{table}
	
		\begin{table*}[h]
		\renewcommand{\arraystretch}{1.1}
		\centering
		\caption{Part of the numerical data of supply (Supply.con)}
		\label{tabela 21}
		\begin{tabular}{|c|c|c|c|c|c|c|c|c|c|c|c|c|c|c|c|c|c|c|c|} 
			\hline 		
			$\#$ & $S_b$  & $P_{S_0}$ & $P_{S}^{max}$  &   $P_{S}^{min}$ &  $P_{S}$ & $C_{P_0}$ & $C_{P_1}$ &  $C_{P_2}$  & $C_{Q_0}$ & $C_{Q_1}$& $C_{Q_2}$&$u_c$&-&$\gamma$&$Q_g^{max}$&$Q_g^{min}$&-&-&u \\
			\hline
			101 &	100	&0.348&	0.348&	0&	0&	0&	0&	0&	0&	0&	0&	0&	0&	1&	0.114&	-0.114&	0&	0&	1\\
			132&	100&	5.04&	5.04&	0&	0&	0&	173.38&	0&	0&	0&	0&	0&	0&	1&	2.440&	-2.440&	0&	0&	1\\			
			\hline 
		\end{tabular} 
	\end{table*}
	\begin{table*}[h]
		\renewcommand{\arraystretch}{1.1}
		\centering
		\caption{Part of the numerical data of demand (Demand.con)}
		\label{tabela 22}
		\begin{tabular}{|c|c|c|c|c|c|c|c|c|c|c|c|c|c|c|c|c|c|} 
			\hline 
			$\#$ & $S_b$  & $P_{D_0}$ & $Q_{D_0}$& $P_{D}^{max}$  &   $P_{D}^{min}$ &  $P_{D}$ & $C_{P_0}$ & $C_{P_1}$ &  $C_{P_2}$  & $C_{Q_0}$ & $C_{Q_1}$& $C_{Q_2}$&$u_c$& -&-&-&u \\
			\hline		
			107 &	100	&0.1208&	0.0172&	0.1208&	1.00E-05&	0&	0&	583.83&	0&	0&	0&	0&	1&	0&	0&	0&	1\\		
			\hline 
		\end{tabular} 
	\end{table*}

		\par This test system adopts as $C_{P_1}$, that is, the linear cost of active power demand, the price of Settlement of Monthly Differences \cite{ASH4:2021:Online}, known as Preço de Liquidação das Diferenças - PLD, which is also determined by CCEE and regulated by ANEEL. The PLD is hourly calculated by each subsystem of the BIPS and determines the prices in the short-term market. The current monthly PLD (July 2021) for the NE subsystem is 583.83\hspace{0.05cm}R\$/MWh (about 111.20\hspace{0.05cm}\$/MWh), which is included in the data structure Demand.con as $C_{P_1}$. The other cost coefficients were left blank as it was not possible to access them, however the data structure allows that the user insert these values.
		\subsection{Part of Supply.con and Demand.con}
		Table~\ref{tabela 21} shows the numerical data of part of Supply.con. 
		The wind farm, \#\hspace{0.05cm}101, present cost coefficients $C_{P_0}$ to $C_{Q_2}$ nulls, and the fossil fuel power plant, \#\hspace{0.05cm}132, whose linear coefficient cost $C_{P_1}$ is 173.38\hspace{0.05cm}R\$/MWh (about 33.00\hspace{0.05cm}\$/MWh), are available for redispatch.
		Table~\ref{tabela 22} shows the numerical data of part of the Demand.con. It is possible to verify that the bus 107-RUSSAS-CE069 has PLD, that is, the  $C_{P_1}$, of 583.83\hspace{0.05cm}R\$/MWh (about 111.20\hspace{0.05cm}\$/MWh).
		\section{Simulation Results}\label{sec:resultados-de-simulacoes}
		This section presents the results from simulations using the test system. These results contribute to evaluate how the system can be used for testing solutions to address the issues of power system with a mix of wind, hydro and fossil fuel power generation, in market and voltage security analysis.  
		\subsection{Simulations of LF, CPF, OPF, and PV curves}
		\par
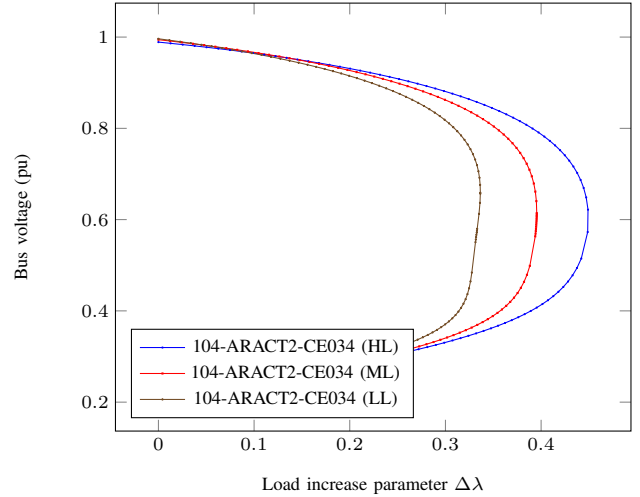
\begin{figure}[h]
			\centering
			\begin{tikzpicture}
				\scriptsize
				\tikzset{every mark/.append style={scale=.1}}
				\begin{axis}[
					xlabel=\scriptsize{{Load increase parameter $\Delta\lambda$}}, 
						ylabel=\scriptsize{Bus voltage (pu)}, 
						legend pos=south west,
						legend entries={
							\scriptsize{104-ARACT2-CE034 (HL)},
							\scriptsize{104-ARACT2-CE034 (ML)},
							\scriptsize{104-ARACT2-CE034 (LL)} }
					]
					\addplot table [x=Loading,y=ARACT2-CE034] {NE_BIPS_229busCoefCustUHEolZero_CurvasPV_FCC_SL_CP3.txt};
					
					\addplot table [x=Loading,y=ARACT2-CE034] {NE_BIPS_229busCoefCustUHEolZero_CurvasPV_FCC_SL_CM3.txt};
					
					\addplot table [x=Loading,y=ARACT2-CE034] {NE_BIPS_229busCoefCustUHEolZero_CurvasPV_FCC_SL_CL3.txt};
				\end{axis}
			\end{tikzpicture}
			\caption[PV curves of critical bus]{PV curves of critical bus}
			\label{FigPV}
		\end{figure}

 Load flow (LF) simulations on the NE-BIPS-229bus, for the base case of the average annual load of the NE subsistem, result in losses of  100.31\hspace{0.05cm}MW. These losses are reduced to 87.51\hspace{0.05cm}MW when the multi-objective optimization is used considering marked and voltage security constraints (Pareto optimal with $\omega = 0.9$). On the other hand, simulation results that can support the study of static voltage stability, presented in Figure \ref{FigPV}, show the PV curves of the critical bus at heavy (HL), medium (ML) and light (LL) loads of the NE subsystem for the month of September 2021. These results, obtained by the use of a CPF, show that the highest voltage stability margin (VSM), $\Delta \lambda\hspace{0.01cm}=\hspace{0.0cm}0.47305$ (1.97\hspace{0.05cm}GW), occurs to the heavy load level.
	
		\subsection{Voltage stability margin sensitivity to wind farm PF}
		The calculation of VSM sensitivities to the wind farms PF results in the ranking shown in Table \ref{RanqueamentodasbarrasPQ}, where the order of the first three wind farms that can best contribute to the VSM is indicated, according to the approach proposal in \cite{silva2019loading}. Thus, the redispatch of these wind farms, with their respective PFs indicated in the table, enable to improve the VSM according to the values of $\Delta \lambda_c$.
	
	\begin{table}[h]
		\renewcommand{\arraystretch}{1.1}
		\centering
		\caption{Rank of wind farm sensitivity}
		\label{RanqueamentodasbarrasPQ}
		\begin{tabular}{|c|c|c|c|}
			\hline
			{\#-Wind farm} &  {Power factor} &  {Sens. $S_i$} &  {$\Delta \lambda_c$} via {CPF}\\ 		 
			\hline
			{101-CQBRD1EOL034} &  {0.95} &  {0.2087}&  {0.4889} \\ 
			{102-BVENTSEOL034} &  {0.95} &  {0.1846}&  {0.4874} \\ 
			{103-ENACELEOL034} &  {0.95} &  {0.1381}&  {0.4827} \\ 
			\hline              
		\end{tabular}
	\end{table}
		\subsection{Optimization considering market and VSC}
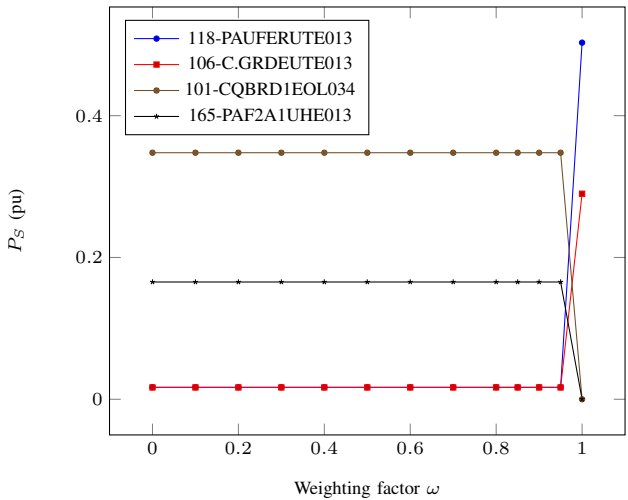
\begin{figure}[h!]
			\centering
			\begin{tikzpicture}
				\scriptsize
				\tikzset{every mark/.append style={scale=.5}}
				\begin{axis}[
					xlabel=\scriptsize{Weighting factor $\omega$},
					ylabel=\scriptsize{$P_{S}$ (pu)},
					legend pos=north west,
					legend entries={\scriptsize{118-PAUFERUTE013},\scriptsize{106-C.GRDEUTE013}, 
						\scriptsize{101-CQBRD1EOL034},\scriptsize{165-PAF2A1UHE013}
					}
					]
					\addplot table [x=weighting,y=PAUFERUTE013] {NE_BIPS_229bus_FPO_Pareto_PS.txt};
					\addplot table [x=weighting,y=C.GRDEUTE013] {NE_BIPS_229bus_FPO_Pareto_PS.txt};
					

					\addplot table [x=weighting,y=CQBRD1EOL034] {NE_BIPS_229bus_FPO_Pareto_PS.txt};
					\addplot table [x=weighting,y=PAF2A1UHE013] {NE_BIPS_229bus_FPO_Pareto_PS.txt};
					
				\end{axis}
			\end{tikzpicture}
			\caption[Power dispatch vs. weighting factor $\omega$]{Power dispatch vs. weighting factor $\omega$}
			\label{FigPSLMCw}
		\end{figure}

		Figure \ref{FigPSLMCw} shows the results of simulations of multi-objective optimization considering market and voltage security constraints for heavy load level. It is possible to see the behavior of the injected power $P_S$ as a function of the the weighting factor $\omega$ in the short-term market, showing that the solution of the multi-objective problem $G$ (\ref{eq:line7}) highlights the transition from a market OPF problem to a voltage security OPF problem, as  $\omega$ is closer to its maximum value 1. For this value, the term that searches for the market optimization presents a null weight. At this point, when the optimization aims only the voltage security, it is possible to see that the fossil fuel power plants (\#\hspace{0.05cm}118 and \#\hspace{0.05cm}106) increase their power dispatch because the generation costs do not present more impact, and the opposite behavior can be seen to the wind farm (\#\hspace{0.05cm}101) and hydro power plant (\#\hspace{0.05cm}165).
		\section{Conclusions}
		This paper presented and described the data structure of the test system NE-BIPS-229bus, which represents part of the NE subsystem. These structures were compiled focusing on the operative conditions of the NE subsystem, and the simulations considered static voltage stability,  voltage stability margin  sensitivities to the wind farms PF, and multi-objective optimization. The results indicate the consistency of the data structure used to represent the test system and its applicability in studies that are intended to analyze the influence of the mix of wind generation with hydro and fossil fuel  power plants. 	
		\par The NE-BIPS-229bus provides a real system, which can be used to validate methodologies and analyses related to the operation of a system with share of wind generation. As future step of this work, in addition to increasing the number of buses, 
		a wide variety of simulations focusing on analyzing of voltage stability, merit order of power plants considering the primary energy source, shunt allocation, grid congestion, contingency analyses N-1, and dispatch optimization of wind and hydro power plants should be performed using the test system. Another perspective for future works, it is the development of a version with the dynamic parameters of the system.
		
		\section*{Acknowledgments}
		The authors thank the São Paulo Research Foundation (FAPESP) for the financial support, number 2018/20104-9.
		\bibliographystyle{unsrt}
		\bibliography{abntex2-modelo-references}

\begin{thebibliography}{10}

\bibitem{ASH3:2021:Online}
ABEEolica. Associacão~Brasileira de~Energia~Eólica.
\newblock Agencia abeeolica, Online:
  http://abeeolica.org.br/agencia-abeeolica/, 2022.
\newblock Access in 02-09-2022.

\bibitem{ASH6:2019:Online}
ONS. Operador~Nacional do~Sistema~Elétrico.
\newblock Mapa dinâmico - sistema de informacões geograficas cadastrais,
  Online: http://www.ons.org.br/paginas/sobre-o-sin/mapas, 2022.
\newblock Access in 02-09-2022.

\bibitem{ASH1:2022:Online}
R.~D. Christie.
\newblock Power systems test case archive, [Online:
  https://labs.ece.uw.edu/pstca/, 1999.
\newblock Access in 09-02-2022.

\bibitem{articleIEEE118up}
I.~Peña, C.~Brancucci Martinez-Anido, and B.~Hodge.
\newblock An extended ieee 118-bus test system with high renewable penetration.
\newblock {\em IEEE Transactions on Power Systems}, PP:1--1, 04 2017.

\bibitem{ASH2:2022:Online}
Washington.
\newblock Sistemas testes brasileiros, Online:
  http://www.sistemas-teste.com.br/, 2011.
\newblock Access in 09-02-2022.

\bibitem{7339813}
A.~Moeini, I.~Kamwa, P.~Brunelle, and G.~Sybille.
\newblock Open data ieee test systems implemented in simpowersystems for
  education and research in power grid dynamics and control.
\newblock In {\em 2015 50th International Universities Power Engineering
  Conference (UPEC)}, pages 1--6, 2015.

\bibitem{7285805}
S.~S. Baghsorkhi.
\newblock Computing saddle-node and limit-induced bifurcation manifolds for
  subtransmission and transmission wind generation.
\newblock In {\em 2015 IEEE Power Energy Society General Meeting}, pages 1--5,
  2015.

\bibitem{6039797}
B.~Tamimi, C.~Cañizares, and K.~Bhattacharya.
\newblock Modeling and performance analysis of large solar photo-voltaic
  generation on voltage stability and inter-area oscillations.
\newblock In {\em 2011 IEEE Power and Energy Society General Meeting}, pages
  1--6, 2011.

\bibitem{574947}
S.~Greene, I.~Dobson, and F.~L. Alvarado.
\newblock Sensitivity of the loading margin to voltage collapse with respect to
  arbitrary parameters.
\newblock {\em IEEE Transactions on Power Systems}, 12(1):262--272, Feb 1997.

\bibitem{hatziargyriou2017contribution}
N.~Hatziargyriou, T.~Van Cutsem, J.~Milanovi{\'c}, P.~Pourbeik, C.~Vournas,
  O.~Vlachokyriakou, P.~Kotsampopoulos, M.~Hong, R.~Ramos, J.~Boemer, et~al.
\newblock Contribution to bulk system control and stability by distributed
  energy resources connected at distribution network.
\newblock 2017.

\bibitem{en9121066}
M.~Abdelrahem and R.~Kennel.
\newblock Fault-ride through strategy for permanent-magnet synchronous
  generators in variable-speed wind turbines.
\newblock {\em Energies}, 9(12), 2016.

\bibitem{meegahapola2013capability}
L.~Meegahapola, T.~Littler, and S.~Perera.
\newblock Capability curve based enhanced reactive power control strategy for
  stability enhancement and network voltage management.
\newblock {\em International Journal of Electrical Power \& Energy Systems},
  52:96--106, 2013.

\bibitem{5159366}
R.~J. Konopinski, P.~Vijayan, and V.~Ajjarapu.
\newblock Extended reactive capability of dfig wind parks for enhanced system
  performance.
\newblock {\em IEEE Transactions on Power Systems}, 24(3):1346--1355, Aug 2009.

\bibitem{5510193}
L.~G. Meegahapola, T.~Littler, and D.~Flynn.
\newblock Decoupled-dfig fault ride-through strategy for enhanced stability
  performance during grid faults.
\newblock {\em IEEE Transactions on Sustainable Energy}, 1(3):152--162, Oct
  2010.

\bibitem{9126776}
M.~Sarkar, T.~Souxes, A.~D. Hansen, P.~E. Sørensen, and C.~D. Vournas.
\newblock Enhanced wind power plant control strategy during stressed voltage
  conditions.
\newblock {\em IEEE Access}, 8:120025--120035, 2020.

\bibitem{8668559}
J.~Ouyang, T.~Tang, J.~Yao, and M.~Li.
\newblock Active voltage control for dfig-based wind farm integrated power
  system by coordinating active and reactive powers under wind speed
  variations.
\newblock {\em IEEE Transactions on Energy Conversion}, 34(3):1504--1511, 2019.

\bibitem{milano2006open}
F.~Milano.
\newblock An open source power system analysis toolbox.
\newblock In {\em Power Engineering Society General Meeting, 2006. IEEE},
  page~1. IEEE, 2006.

\bibitem{ASH:20214:Online}
V.~Neumann, R.~Kuiava, R.~Ramos, and A.~Piovani.
\newblock Ne-bips-229bus test system, Online:
  https://drive.google.com/drive/folders/1O32J-3xkd-gBcZMXw8snZg6aan8JsGko?usp=sharing,
  2022.
\newblock Accesss in 02-07-2022.

\bibitem{4334890}
Working~Group on~Exchange~of Power System Analytical~Data.
\newblock Proposed data structure for exchange of power system analytical data.
\newblock {\em IEEE Transactions on Power Systems}, 1(2):8--15, 1986.

\bibitem{ASH9:2017:Online}
ONS. Operador~Nacional do~Sistema~Elétrico.
\newblock Procedimentos de rede – submódulo 5.6, Online:
  www.ons.org.br/Paginas/busca.aspx?k=procedimentos\%20de\%20rede, 2017.
\newblock Access in 07-15-2021.

\bibitem{ASH8:2019:Online}
MME-CEPEL.~Ministério de~Minas~e Energia.
\newblock Representação dos patamares de carga na cadeia de modelos
  computacionais do setor elétrico, Online:
  www.mme.gov.br/documents/10584/61476853/Ata+Anexo+3+CP
  +51+Relat
  - 2019.
\newblock Access in 07-15-2021.

\bibitem{ASH:2021:Online}
EPE.~Empresa de~Pesquisas~Energéticas.
\newblock Plano decenal de energia 2030, Online:
  https://www.epe.gov.br/sites-pt/publicacoes-dados-abertos/publicacoes/PublicacoesArquivos/publicacao-490/PDE
  2021.
\newblock Access in 02-09-2022.

\bibitem{667386}
L.~M. Hajagos and B.~Danai.
\newblock Laboratory measurements and models of modern loads and their effect
  on voltage stability studies.
\newblock {\em IEEE Transactions on Power Systems}, 13(2):584--592, May 1998.

\bibitem{ASH2:2021:Online}
EPE.~Empresa de~Pesquisas~Energéticas.
\newblock Modelo de mercado da micro e minigeração distribuída (4md):
  Metodologia, versão pde 2031, Online:
  https://www.epe.gov.br/sites-pt/publicacoes-dados-abertos/publicacoes/PublicacoesArquivos/publicacao-490/topico-531/,
  2022.
\newblock Access in 02-09-2022.

\bibitem{5994933}
J.~Jung and W.~Hofmann.
\newblock Investigation of thermal stress in the rotor of doubly-fed induction
  generators at synchronous operating point.
\newblock In {\em 2011 IEEE International Electric Machines Drives Conference
  (IEMDC)}, pages 896--901, May 2011.

\bibitem{7232809}
J.~A. Martin and I.~A. Hiskens.
\newblock Reactive power limitation due to wind-farm collector networks.
\newblock In {\em 2015 IEEE Eindhoven PowerTech}, pages 1--6, June 2015.

\bibitem{ASH:20216:Online}
ANEEL. Agencia~Nacional de~Energia~Elétrica.
\newblock Validador eol - sisgel, [Online:]
  https://sigel.aneel.gov.br/validadoreol/, 2021.
\newblock Access in 07-15-2021.

\bibitem{milano2005voltage}
F.~Milano, C.~A. Canizares, and M.~Invernizzi.
\newblock Voltage stability constrained opf market models considering n- 1
  contingency criteria.
\newblock {\em Electric Power Systems Research}, 74(1):27--36, 2005.

\bibitem{1198291}
F.~Milano, C.A. Canizares, and M.~Invernizzi.
\newblock Multiobjective optimization for pricing system security in
  electricity markets.
\newblock {\em IEEE Transactions on Power Systems}, 18(2):596--604, 2003.

\bibitem{monticelli1983fluxo}
A.~J. Monticelli.
\newblock {\em Fluxo de carga em redes de energia el{\'e}trica}.
\newblock E. Blucher, 1983.

\bibitem{9638214}
V.~Neumann, R.~Kuiava, R.~Ramos, A.~Piovani, and L.~Alburguetti.
\newblock Loading margin improvement through adjustement of the wind farms
  power factor.
\newblock In {\em 2021 IEEE Power Energy Society General Meeting (PESGM)},
  pages 1--5, 2021.

\bibitem{silva2019loading}
V.~Neumann and R.~Kuiava.
\newblock Loading margin sensitivity in relation to the wind farm generation
  power factor for voltage preventive control.
\newblock {\em Journal of Control, Automation and Electrical Systems},
  30(6):1041--1050, 2019.

\bibitem{ASH4:2021:Online}
CCEE.~Câmara de~Comercio~de Energia~Elétrica.
\newblock Preço de liquidação das diferenças - pld, Online:
  https://www.ccee.org.br/portal/, 2021.
\newblock Access in 07-15-2021.

\end{thebibliography}
	\end{document}